\documentclass[cup6b]{cupbook}

\usepackage{epsfig}
\usepackage{graphics}
\usepackage{bm}

\newcommand{\Eq}[1]{Eq.~(\ref{#1})}
\newcommand{\be}{\begin{equation}}
\newcommand{\ee}{\end{equation}}

\begin{document}

\pagenumbering{arabic}
\author[J. D. Bekenstein]{Jacob D. Bekenstein\\Racah Institute of Physics\\Hebrew University of Jerusalem}
\chapter[Alternatives to dark matter]{Modified gravity as an alternative to dark matter}

\hyphenation{}
\begin{abstract}
The premier alternative to the dark matter paradigm is modified gravity.  Following an introduction to the relevant phenomenology of galaxies, I review the MOND paradigm, an effective summary of the observations which any theory must reproduce.  A simple nonlinear modified gravity theory does justice to MOND at the nonrelativistic level, but  cannot be elevated to the relativistic level in a unique way.  I go in detail into the covariant tensor-vector-theory (T$e$V$e\,$S) which not only recovers MOND but can also deal in detail with gravitational lensing and cosmology.  Problems with MOND and T$e$V$e\,$S at the level of clusters of galaxies are given attention.  I also summarize the status of T$e$V$e\,$S cosmology.
\end{abstract}

\section*{1. Introduction}\label{intro}

A look at the other papers in this volume will show the present one to be singular.  Dark matter is a prevalent paradigm.  So why do we need to discuss alternatives ?  While observations seem to suggest that disk galaxies are embedded in giant halos of dark matter (DM), this is just an \textit{inference} from accepted Newtonian gravitational theory.  Thus if we are missing understanding about gravity on galactic scales, the mentioned inference may be deeply flawed.  And then we must remember that, aside for some reports which always seem to contradict established bounds, DM is not seen directly.  Finally, were we to put all our hope on the DM paradigm, we would be ignoring a great lesson from the history of science: accepted understanding of a phenomenon has usually come through confrontation of rather contrasting paradigms.

\begin{figure}
      \centering
    \includegraphics[width=10cm]{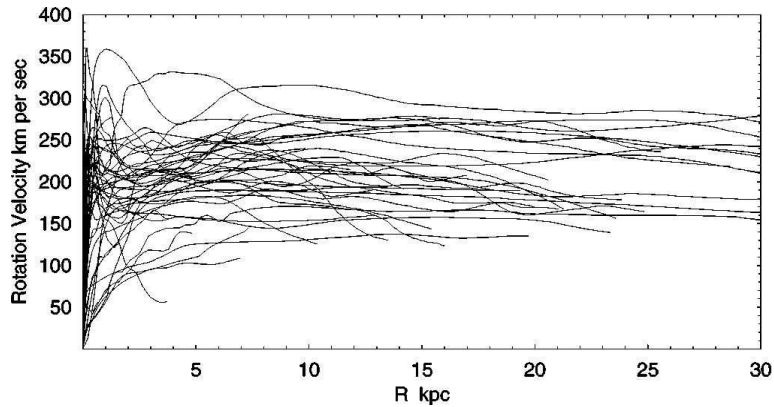}
   \caption{Examples of rotation curves of nearby spiral galaxies from Sofue and Rubin (2001).  These  resulted from combining Doppler data from CO molecular lines for the central regions, optical lines for the disks, and HI 21 cm line for the outer gaseous regions.  The galactocentric radius $R$ is in kiloparsec (1 kpc $\approx$ 3000 light years).   Graph reprinted, with permission, from the Annual Review of Astronomy and Astrophysics, Volume 31 (c)2001 by Annual Reviews
www.annualreviews.org.}   
 \label{figure:fig1}
 \end{figure}

To construct a competing paradigm to DM, it is best to bear in mind concrete empirical facts.  Newtonian gravity with the visible matter as source of the Poisson equation properly describes all  observed systems from asteroid scale up to the scale of the globular clusters of stars ($\sim 10^5$ stars bound together in a ball the size of a few tens of light years).  But as we move up to galaxies, ours or external ones, troubles appear.  In essence the way disk-like galaxies rotate is incompatible with the Newtonian gravitational force generated by only the visible stars, gas and dust.
From the centrally concentrated light distribution of the typical disk galaxy we would expect a rotation linear velocity which first rises with galactocentric radius $r$ and then drops asymptotically as $r^{-1/2}$.  But as clear from Fig.\ref{figure:fig1}, most disk galaxy rotation curves become flat and stay so to the outermost measured point, which generally lies well beyond the edge of the optical galaxy.  And typically the mass lying within the last measured point of the rotation curve, as calculated \textit{a la Newton}, is at least an order of magnitude larger than the baryonic mass actually seen.  These are empirical facts begging explanation.  Dark matter, if appropriately distributed in each case, can explain the shape and scale of rotation curves.  But the required mass distributions are not always reasonable from the point of view of galactogenesis. e.g., the predicted central cusps in density are not observed.

\begin{figure}
\centering
\includegraphics[width=5cm]{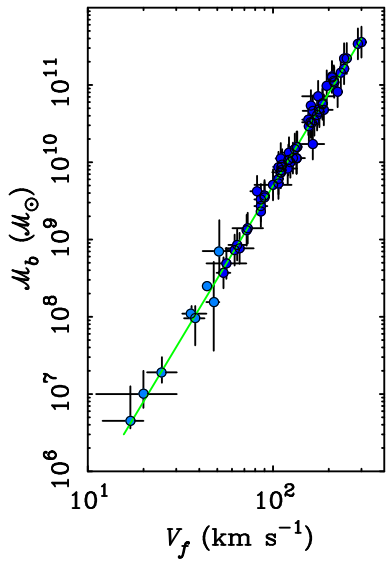}
  \caption{The baryonic Tully-Fisher correlation from McGaugh (2005) for disk galaxies. $\mathcal{M}_b$ is the total baryonic (stars plus gas) mass; $V_f$ is the asymptotic rotation velocity.     For  galaxies from Sanders and McGaugh (2002)  (shaded circles) the mass in stars comes from a fit of the shape  of the rotation curve with MOND.  For the eight dwarf spirals (unshaded circles) the mass in stars (relatively small)  is inferred directly from the luminosity.  The solid line, with slope 4, is MOND's prediction.  Graph reproduced  by permission of the American Astronomical Society.}
 \label{figure:fig2}
 \end{figure}

Another fact to be explained is the Tully-Fisher law for disc galaxies.  Originally discovered as a correlation between blue luminosity and the peak rotation velocity of a disk, it has metamorphosed into McGaugh's baryonic Tully-Fisher law: \textit{the total mass in visible stars and gas (baryonic mass) in a disc galaxy is accurately proportional to the fourth power of the asymptotic  (terminal) rotational velocity of that galaxy}.  This law extends over six orders of magnitude in mass, and is a tight correlation as can be appreciated from Fig.\ref{figure:fig2}.   
In the low surface brightness galaxies (LSBs), those with low central brightness per unit area (typically below 21.65 magnitudes per square arcsec), the rotation curve it still on the rise at the last measured point.   Yet LSBs, which can be luminous and massive or small dim galaxies, all fall on the same Tully-Fisher relation as more conventional disks (McGaugh 2005).  
 
Within the DM paradigm the  Tully-Fisher law must arise from galaxy formation since it connects luminosity of baryonic matter with a dynamical property, rotation, which is seen as  dominated by the DM halo.  But it has not been easy to derive Tully-Fisher from any natural connection between the two components.   And as R. H. Sanders has pointed out, the messiness of galaxy formation is hardly the natural backdrop for such a sharp correlation between galaxy properties.  The sharpness needs a dynamical reason as opposed to an evolutionary one.

Turn from galaxies to clusters of galaxies.  In these systems, containing sometimes hundreds of galaxies and much hot intergalactic gas, the Newtonian virial theorem can be used to estimate the cluster's mass from the velocities of galaxies in the cluster.  The determined masses are very large  compared to the mass seen directly as galaxies and hot gas.   The mass discrepancy also shows up when the overall cluster mass is determined \textit{a la Newton} from the assumption that the hot gas is in a hydrostatic state, or when gravitational lensing by a cluster is analyzed in the framework of general relativity (GR).  The conventional solution is to assume that the typical cluster contains DM to the tune of about five times the visible mass.

There are other aspects of the missing mass problem, but the above will furnish enough background for the ensuing discussion here.  The questions before us are two.  Are there other scenarios, apart from DM, which can account for all the mentioned facts.   And can one single out a particular alternative scenario as especially promising, both in terms of physical basis, and explanatory success? 

\section*{2. The MOND scheme}\label{MOND}

\begin{figure}
\centering
\includegraphics[width=6cm]{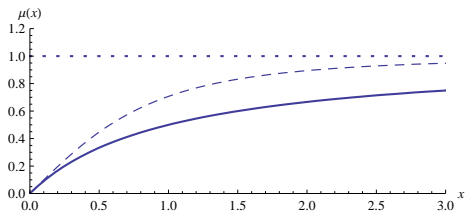}
  \caption{Two widespread choices  for Milgrom's interpolation function: the ``simple'' function $\tilde\mu(x)=x/(1+x)$ (solid) and the  ``standard'' function $\tilde\mu(x)=x/\sqrt{1+x^2}$ (dashed).  The dotted line corresponds to strict Newtonian behavior, obtainable also as the limit $a_0\rightarrow 0$ of MOND.}
 \label{figure:fig3}
 \end{figure}

The MOND scheme or paradigm of  Milgrom(1983a,b,c) serves first and foremost as an effective summary of much extragalactic phenomenology.  Milgrom introduced a preferred scale of acceleration $a_0\approx 10^{-10}\,  {\rm m\, s}^{-2}$ of the order of the centripetal accelerations of stars and gas clouds in the outskirts of disk galaxies.  In terms of it MOND relates the acceleration ${\bm a}$ of a test particle to the ambient Newtonian gravitational field $-\bm\nabla\Phi_N$ generated by the baryonic mass density alone by
\begin{equation}
\tilde\mu(|{\bm a}|/a_0)\,{\bm a}=-\bm\nabla\Phi_N.
\label{MOND}
\end{equation}
Milgrom assumes that the positive smooth monotonic function $\tilde\mu$ approximately equals its argument when this is small compared to unity (deep MOND regime), but tends to unity when that argument is large compared to unity (Fig.\ref{figure:fig3}).   

The stars or gas clouds orbiting in the disk of a spiral galaxy of baryonic mass $M$ at radius $r$ from its center do so in nearly circular motion with velocity $V(r)$.  Obviously $|{\bm a}|=V(r)^2/r$.  Sufficiently outside the main mass distribution we may estimate $|\bm\nabla\Phi_N|\approx GM/r^2$.  And at sufficiently large $r$, $|{\bm a}|$ will drop below $a_0$ and we should be able to approximate $\mu(x)\approx x$.  Putting all this together gives $V(r)^4/r^2\approx  G M a_0/r^2$.  It follows, first,  that  well outside the main mass distribution $V(r)$ must become independent of $r$, that is, the rotation curve must flatten at some value $V_f$, in agreement with observations.  Second, from the coefficients it follows that
\begin{equation}
M= (G a_0)^{-1}V_f{}^4.
\label{TFlaw}
\end{equation}
But this is just McGaugh's baryonic Tully-Fisher law (Fig.~\ref{figure:fig2}).   The MOND predicted proportionality coefficient $(G a_0)^{-1}$ agrees well with the measured coefficient of the baryonic Tully-Fisher law, namely  $50\, M_\odot\, {\rm km}^{-4}\,s^{-4}$.  Thus, MOND's single formula unifies two central facts of spiral galaxy phenomenology. 

\begin{figure}
\centering
\includegraphics[width=11.5cm]{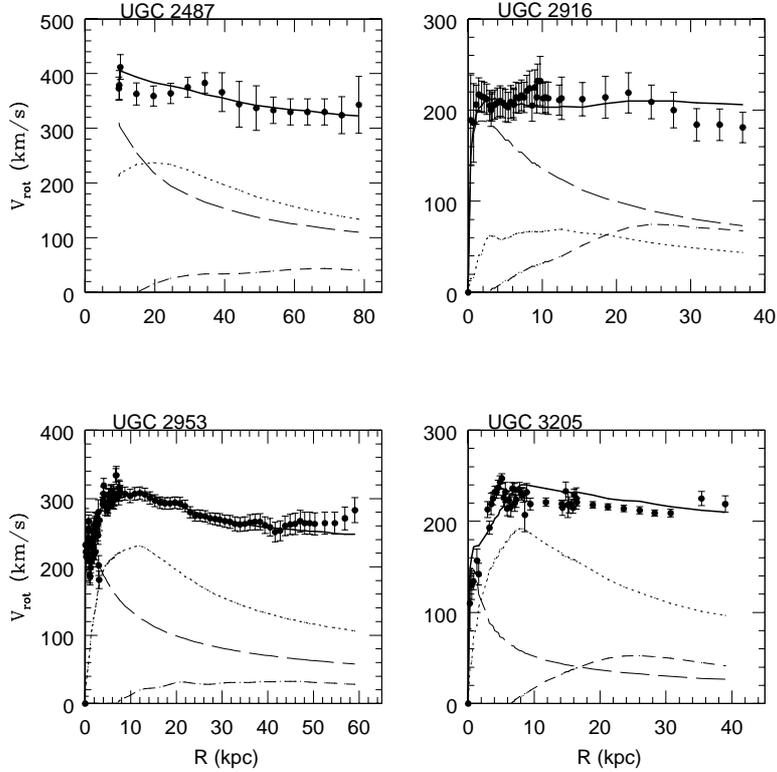}
  \caption{MOND fits to the measured rotation curves of four disk galaxies from the Uppsala catalog modeled by the method of  Sanders and Noordermeer (2007).   Each solid curve is the fit  (based on the simple function $\tilde\mu$ of Fig.~\ref{figure:fig3})  generated by the distribution of stars and neutral hydrogen.   Long-dashed, dotted and short-dashed curves give the Newtonian rotation curves generated separately  by the galaxy's bulge, disc and gas, respectively.  Figure courtesy of R. H. Sanders.}
 \label{figure:fig4}
 \end{figure}

Of course a devil's advocate could claim that MOND works well for the above phenomena because the low argument form of $\tilde\mu(x)$ and the value of $a_0$ have been rigged to obtain these results: after all, both flat rotation curves and early forms of the Tully-Fisher law were already known before MOND was formulated.  Note, however, that  the supposed prearrangement need not guarantee  that the parameter $a_0$ needed to recover the observed  Tully-Fisher law should have anything to do with the detailed shape of rotation curves interior to the flat regions.  Yet MOND is singularly successful in explaining the detailed \emph{shapes} of rotation curves, as made clear by Fig.\ref{figure:fig4}.  The observed points are radio determined velocities.  The MOND predictions were made using the ``simple'' form of $\tilde\mu(x)$.  Once the value of $a_0$ is adopted, the only parameter that need be adjusted to fit the velocities using the observed  photometry is the stellar mass-to-luminosity ratio $\Upsilon$ of each disk.  The required $\Upsilon$s turn out, in most cases, to be reasonable in view of stellar evolutionary models (Sanders and McGaugh 2002).   So MOND is a successful and consistent one-parameter-per-galaxy paradigm.  By contrast DM halo models usually require three fitting parameters, e.g. $\Upsilon$, length scale, and central velocity dispersion, and yet they do not do a better job than MOND.  For disk galaxies MOND is more economical, and more falsifiable, than the DM paradigm. 

One cannot emphasize enough that \textit{a priori} the $a_0$ that enters in the Tully-Fisher law is, in principle, a different parameter from that which determines the shape of rotation curves. The value $a_0= 1.2\cdot 10^{-10}\,  {\rm m\, s}^{-2}$ that needs to be adopted for good fits to over one hundred  galaxies  agrees very well with the $a_0$ determined from the baryonic Tully-Fisher law.    
 It is only in the MOND paradigm that the two roles of $a_0$ have a common origin, and observations agree that there is only one $a_0$.   

Yet a third role for $a_0$ results from Milgrom's observation that $\Sigma_m=a_0/G$ sets a special scale for mass surface density.   Wherever in a system the actual surface mass density drops below $\Sigma_m$, Newtonian gravitational behavior gives way to MOND dynamics (Milgrom 1983b). In galaxies like ours this occurs some way out in the disk; that is why the rotation curves may exhibit a brief drop (attempting to act \textit{a la Newton}) before becoming flat asymptotically as befits MOND.  Of course, if at the center of a disk the surface mass density is already below $\Sigma_m$, then MOND holds sway over the whole disk. Thus Milgrom predicted that were disk galaxies to exist with surface mass densities everywhere below $\Sigma_m$, they should show especially large acceleration discrepancies coming from the fact that $\tilde\mu\ll1$ everywhere in them.   Likewise he noted that the shapes of the rotation curves should be independent of the precise way $\tilde\mu(x)$ changes from a linear function to a constant (unity) since the nonlinear range of $\mu$ does not come into play.    A population of such galaxies became known in the late 1980's (Begeman et al. 1991); these are none other than the LSBs. Much subsequent work has confirmed both facets of Milgrom's prediction (McGaugh and de Blok 1998).  In particular, it is easy to understand why in these galaxies the rotation curves are still on the rise at the outermost observed point.

Use of the MOND analog to the virial theorem (Milgrom 1994a)  to estimate the masses of clusters softens, but does not fully resolve, the acceleration discrepancy. Clusters still seem to contain a factor of 2-3 more matter than is actually observed in all known forms (Sanders and McGaugh 2002).  One should keep in mind that clusters may contain much invisible matter of rather prosaic nature, either baryons in a form which is hard to detect optically, or massive neutrinos (Sanders 2003, 2007; Pointecouteau 2006) so that, while troubling, the above lingering discrepancy should not stymie further investigation of MOND.

Excellent reviews exist of other phenomenological successes of MOND (Sanders and McGaugh 2002, McGaugh 2006, Scarpa 2006, Milgrom 2008).    So I forego further review of MOND's implications, and turn to ask the obvious question.

\section*{3. Modified gravity theory for MOND}\label{vac}

There is no doubt that MOND is a useful paradigm for summarizing extragalactic data.  But what is the physical basis of its success?  The most conservative answer is that MOND is merely a happy summary of how much DM there is in galaxies, and how it is spatially distributed in the various types of objects.    Surprisingly, this minimal interpretation, which would surely excite no opposition from conventional astrophysicists, \textit{does not} hold out well.  First, such interpretation should explain how the scale $a_0$ enters into spiral galaxy properties through some regularity in the formation process of DM halos.  Yet detailed arguments have shown that the halos cannot have an intrinsic scale (Sanders and Begeman 1994).  In addition,  it is hard to see how such an interpretation could give an account of the multiple roles of $a_0$ alluded to above (Milgrom 2002). 

Another interpretation of MOND (Milgrom 1983b,1994b) is that it reflects modified inertia, e.g. \Eq{MOND} represents a modification of the Newtonian law $\bm{f}=m\bm{a}$, whether the force $\bm{f}$ is purely gravitational or includes other forces.  In evaluating this alternative it is important to realize that the MOND formula cannot be exact.  For in a binary system with unequal masses, $m_1$ and $m_2$, the time derivative of $m_1{\bm v}_1+m_2{\bm v}_2$, as calculated from \Eq{MOND}, does not vanish in general since the inertial factors $\tilde\mu$'s will generally be unequal.  In light of this failure of momentum conservation,  Milgrom suggested that \Eq{MOND} is meaningful only for  test particle motion in a background gravity field.  While this restriction is not onerous if all one  wants is to analyze the rotation curve of a galaxy,  it would put problems such as the dynamics of a binary galaxy system out of the reach of MOND.

Even as a theory of test particle motion, the modified inertia interpretation of \Eq{MOND} is problematic.   Milgrom (1994b) showed rigorously that no Lagrangian exists for it which (a) is Galilean invariant, (b) generates Newtonian dynamics for $a_0\rightarrow 0$ and (c) generates deep MOND dynamics when $a_0\rightarrow \infty$.  (If one instead demands relativistic invariance, the desired Lagrangian is even more remote (Domokos 1996)).  To break this \textit{impasse} Milgrom suggested giving up the locality requirement and instead deriving the modified inertia from a nonlocal action functional, one which is not an integral over a Lagrangian density.  More recently Milgrom (2006) has suggested, as an alternative route, to give up Galilean invariance in order to make a Lagrangian possible.  Clearly the modified inertia viewpoint of MOND is a difficult one to formalize.

When introducing MOND Milgrom also suggested that it may instead represent a modification of the Newtonian law of gravitation.   To see this just replace the r.h.s. of $m{\bm a}={\bm f}$, with a non-Newtonian gravitational force which depends nonlinearly on its sources in such a way that the content of this equation coincides that of \Eq{MOND}.  Of course, the resulting theory will also flout the conservation of momentum for the same reason as above.  This defect is easily repaired by deriving the requisite theory from a suitable Lagrangian (Bekenstein and Milgrom 1984).  

First one stipulates that a particle's acceleration ${\bm a}$ is given by $-{\bm \nabla}\Phi$ where $\Phi$ is some potential distinct from Newton's; in this way the motion is conservative.  As the Lagrangian for $\Phi$ one selects the simplest generalization of the Newtonian field Lagrangian which is still rotationally and Galilean invariant, but depends on a scale of acceleration $a_0$ which should be identified with MOND's:
\begin{equation}
L = -\int\Big[{a_0{}^2\over 8\pi G} F\Big({|\bm\nabla\Phi|^2\over a_0{}^2}\Big) + \rho\Phi\Big]d^3x.
\label{lagrangian}
\end{equation}
Here  $F$ is some positive function and $\rho$ is the visible matter's mass density.   This Aquadratic Lagrangian theory (AQUAL) reduces to Newton's for the choice $F(X^2)= X^2$, when it readily leads to Poisson's equation for $\Phi$.  Generically it yields the equation
\begin{eqnarray}
\bm\nabla\cdot[\tilde\mu(|\bm\nabla\Phi|/a_0)\bm\nabla\Phi]&=&4\pi G\rho,
\label{AQUAL}
\\
\tilde\mu(X) &\equiv& dF(X^2)/d(X^2).
\label{F}
\end{eqnarray}
The $\tilde\mu$ function here should be identified with Milgrom's $\tilde\mu$ for the following reason.

Comparison of the AQUAL equation,~\Eq{AQUAL}, with Poisson's for the same $\rho$ allows to determine its first integral,
\begin{equation}
\tilde\mu(|\bm\nabla\Phi|/a_0)\bm\nabla\Phi=\bm\nabla\Phi_N+\bm\nabla\times {\bm  h},
\label{integral}
\end{equation}
where ${\bm h}$ is some calculable vector field.   Now in situations with spherical or cylindrical symmetry ${\bm h}$ must vanish by symmetry, and it is plain to see that \Eq{integral} reduces to the MOND equation because ${\bm a}=-{\bm \nabla}\Phi$.  Thus, at least for highly symmetric situations, the MOND equation can be recovered exactly; in more realistic geometries it acquires some ``corrections''.
AQUAL has the correct limits if we stipulate that $F(X^2)\rightarrow X^2$ for $|{\bm \nabla}\Phi|\gg a_0$ so that $\tilde\mu$ is constant (Newtonian regime), and $F(X^2)\rightarrow {\scriptstyle 2\over \scriptstyle 3} X^{3/2}$ for $|{\bm \nabla}\Phi|\ll a_0$ so that $\tilde\mu(|\bm\nabla\Phi|/a_0) \rightarrow |\bm\nabla\Phi|/a_0$ in this deep MOND regime.

At present AQUAL is the best available embodiment of the MOND idea in the nonrelativistic regime.  Being Lagrangian based, AQUAL respects the energy, momentum and angular momentum conservation laws.  It removes an ambiguity inherent in the MOND formula: how is a composite object, e.g. a star, to move if its constituents (ions) are subject to strong collision related accelerations  ($\gg a_0$), but the composite moves as a whole in a weak  gravitational field ($\ll a_0$), e.g. the far field of the Galaxy?  AQUAL makes it clear that, regardless of the internal makeup, the composite's center of mass moves in the ambient field according to the MOND formula (Bekenstein and Milgrom 1984).  This is in harmony with the fact that all sorts of stars, gas clouds, etc. delineate one and the same rotation curve for a galaxy.  AQUAL has permitted the calculation of the MOND force between two galaxies in a binary from first principles (Milgrom 1986, 1994a).  It establishes the existence of the conjectured ``external field effect'', namely, the partial suppression of MOND effects in a system with internal accelerations weak on scale $a_0$ which happens to be immersed in a gravitational field strong on scale $a_0$ (Bekenstein and Milgrom 1984).  This may well explain the total absence of missing mass effects in the open galactic clusters, systems with weak internal accelerations (Milgrom 1983a). 

Newtonian disk instabilities originally motivated the idea that DM halos surround disk galaxies, and by dint of their dominant gravity were thought to moderate the dangerous instabilities.   Using a mixture of MOND formula and AQUAL equation arguments, Milgrom (1989) showed  analytically that MOND enhances local stability of disks against perturbations as compared to the Newtonian situation.
And AQUAL has opened the door to numerical simulations of stellar systems within the MOND paradigm (Milgrom 1986) which help to explore complex phenomena like stability.  Thus Brada and Milgrom (1999) combined a $N$-body code with a numerical solver for the AQUAL equation to verify that MOND stabilizes galaxy discs locally,  and to demonstrate that it enhances global stability as well.  The degree of stabilization saturates in the deep MOND regime so no disk is absolutely stable in MOND.  This agrees with the fact that a fraction of disk galaxies have bars which must have arisen from instabilities.  

More recently the Paris group of Tiret and Combes  studied \emph{evolution} of bars in spiral disks using their own  AQUAL based $N$-body code.  They find that bars actually form more rapidly in MOND than in DM halo Newtonian models, but that MOND bars then weaken as compared to those in Newtonian models.  The long term distribution of bar strengths predicted by MOND is in better agreement with the observed distribution than that obtained from Newtonian halo models (Tiret and Combes  2007, 2008).  Another N-body simulator based on AQUAL has been developed by the Bologna school of Ciotti, Nipoti and Londrillo, and used  in a number of investigations.  In studying dissipationless collapse according to AQUAL, they find that the profiles of the ensuing objects  look like those of real elliptical galaxies, and when the resulting dynamics are  interpreted Newtonially, the objects seem to contain little DM, as indeed found in real life (Nipoti et al. 2007a).  In their study of merging galaxies with AQUAL  they find that merging is slower than that in Newtonian theory  (because of the lack of DM generated dynamical friction).  This may be a problem for MOND because of widespread evidence that galaxies merge fast.  However, the outcome of a merger  is indistinguishable from the merger product in Newtonian theory with DM (Nipoti et al. 2007b).  In joint work with Binney (Nipoti et al. 2008) the Bologna group have investigated dynamical friction with the AQUAL based code confirming the conclusion of an earlier analytic study (Ciotti and Binney 2005) that MOND shortens the dynamical friction timescale as compared to that of a system with the same phase space distribution in Newtonian theory with DM.  The situation with regard to slowing of bars and mergers is satisfactory; however, a persistent problem remains with regard to the too brief timescale for digestion of globular clusters in the dwarf spheroidal galaxies (Ciotti and Binney 2004,  S\'anchez-Salcedo et al.  2006).

As could be expected, AQUAL, being nothing but a reformulation of the MOND equation, does not resolve the acceleration discrepancy in clusters of galaxies.  We shall return to this at the end. 
For a more comprehensive review of the AQUAL representation of MOND, the reader is referred elsewhere (Bekenstein 2006).

\section*{4. T$e$V$e\,$S and other relativistic MOND theories}

Neither the MOND formula nor AQUAL pretend to encompass relativistic phenomena.  Yet we need to deal with gravitational lenses, gravitational waves, and cosmology, all relativistic phenomena connected with DM.  How to make the MOND idea relativistic?  Since AQUAL has had significant success, we should start with it.  

The obvious way to make AQUAL into a covariant theory is to let the AQUAL potential $\Phi$ metamorphose into a scalar field $\phi$ whose Lagrangian is the covariant version of \Eq{lagrangian}.   Einstein's equations of GR are retained so that the desired theory will approximate standard gravity theory, and so partake of the latter's many successes, e.g. in the solar system.  To achieve the requisite deviation from exact GR, which after all cannot reduce to MOND, one couples matter  in the new theory to $\phi$ by the simple technique of writing the \textit{matter} Lagrangian entirely with the new metric $\tilde g_{\alpha\beta}=e^{2\phi}\, g_{\alpha\beta}$, where henceforth we set $c=1$. This program was carried out (Bekenstein and Milgrom 1984) but the resulting relativistic AQUAL turns out to have superluminal propagation (of $\phi$ waves).  Even more disconcerting is its failure to deal with the observational fact that, whatever it is that enhances the pull of galaxies and clusters on their components, also significantly enhances their gravitational lensing power (Bekenstein 1988).  The root of this problem is the conformal relation between the two metrics:  it is well known that conformally related metrics are equivalent insofar as light propagation is concerned. 

It took a while to gather together the elements for a viable relativistic embodiment of MOND.  One of these is a timelike 4-vector field with unit norm, $\mathcal{ U}_\alpha$, first proposed by Sanders (1997). It is needed to break the above mentioned conformal relation, which Sanders replaced by
\begin{equation}
 \tilde g_{\alpha\beta}=e^{-2\phi}\,  g_{\alpha\beta}- (e^{2\phi}-e^{-2\phi})\, \mathcal{ U}_\alpha\,\mathcal{ U}_\beta.
 \label{gtilde}
 \end{equation}
The spacetime described by $\tilde g_{\alpha\beta}$ is related to that of $g_{\alpha\beta}$ by a stretching in the direction of $\mathcal{U}^\alpha\equiv g^{\alpha\beta}\mathcal{ U}_\beta$ and a compression orthogonal to it.  
 Sanders regarded  $\mathcal{ U}^\alpha$ as established by the cosmos, and as pointing in the time direction. 
 With this proviso, the problem with the lensing can be cured.  
 
 However, the prescription that a 4-vector point in the time direction is not a covariant one.  Instead one should endow $\mathcal{ U}_\alpha$ with covariant dynamics of its own.  In the  tensor-vector scalar theory, or T$e$V$e\,$S (Bekenstein 2004a and 2004b, Bekenstein and Sanders 2006), these dynamics are derived from a covariant action of the general form ($K$ and $\bar K$ are dimensionless coupling constants)
\begin{eqnarray}
S_v &=&-{ 1\over 32\pi G}\int
\Big[K g^{\alpha\beta}g^{\mu\nu}  \mathcal{ U}_{[\alpha,\mu]} \mathcal{ U}_{[\beta,\nu]} + \bar K ( g^{\alpha\beta}  \mathcal{ U}_{\alpha;\beta})^2
\\
&-&2 \lambda(g^{\mu\nu}\mathcal{ U}_\mu
\mathcal{ U}_\nu +1)\Big](-g)^{1/2} d^4 x.
\end{eqnarray}
The first term in the integrand is a Maxwell-like one; it takes care of approximately aligning $\ \mathcal{ U}^\alpha$ with the 4-velocity of matter in the region in question, this in lieu of the noncovariant requirement that  $\ \mathcal{ U}^\alpha$ point in the time direction.  But the ``gauge freedom'' inherent in this term turns out to cause integral lines of $\mathcal{ U}_\alpha$ to form caustics (Contaldi et al. 2008).  Thus in later work the second term in the integrand is introduced to break the gauge freedom.    Finally, the last term is a constraint ($\lambda$ being a Lagrange multiplier function) which forces $\mathcal{ U}_\alpha$ to have unit negative norm with respect to metric $g_{\alpha\beta}$, and so to always be  timelike.  Zlosnik, et al. (2006) have shown how to reformulate T$e$V$e\,$S  solely in terms of the metric $\tilde g_{\alpha\beta}$ (which turns out to satisfy Einstein-like equations), and the vector field; the $\phi$ is eliminated with the help of the above constraint.   In this form  T$e$V$e\,$S  might figuratively be described as GR with a DM (the vector field).  However, the second form is very intricate.    We continue to discuss T$e$V$e\,$S in its original form.

The dynamics of $\phi$ in relativistic AQUAL must be modified to obviate the superluminal propagation (Bekenstein 2004a).  Here we use the notation of later work (Sagi and Bekenstein 2008) to write the scalar action as
\begin{equation}
\label{scalaraction}S_s=-\frac{1}{2 k^2 \ell^2 G}\int
\mathcal{F}\left(k
\ell^2h^{\alpha\beta}\phi_{,\,\alpha}\phi_{,\,\beta}\right)\,\sqrt{-g}\,d^4x,
\end{equation}
where $\mathcal{F}$ is a positive function, $h^{\alpha\beta}\equiv g^{\alpha\beta}-\mathcal{ U}^\alpha\mathcal{ U}^\beta$, $k$ is another dimensionless coupling constant, and $\ell$ is a constant scale of length.  Each choice of the function $\mathcal{F}(y)$ defines a separate TeVeS
theory; $\mu(y)=d\mathcal{F}(y)/dy$ functions somewhat  like the $\tilde{\mu}(x)$  in MOND.   We
need only consider functions such that $\mathcal{F}>0$ and $\mu>0$
for either positive or negative arguments.   The correct MOND behavior will emerge if $\mu(y)\rightarrow 1$ for $y\rightarrow\infty$ and $\mu(y)\approx D \surd y$
for $0<y\ll 1$  with $D$ a dimensionless positive constant. 
The action of T$e$V$e\,$S comprises $S_v$, $S_s$ as well as the customary Einstein-Hilbert action for $g_{\alpha\beta}$ and the matter action written in the usual fashion entirely in terms of $\tilde g_{\alpha\beta}$.
 
Consequences of T$e$V$e\,$S with the above  sort of $\mathcal{F}$ have been investigated mostly for the case $\bar K=0$.  T$e$V$e\,$S reduces to GR in the parameter regime $k\to 0$, $K\propto k$ and $\ell\propto k^{-3/2}$ (Bekenstein 2004a).  As a consequence there is a nonrelativistic quasistationary regime in which T$e$V$e\,$S is Newtonian, and which serves to describe the overall features of Earth and solar system gravity.  In contrast to many gravity theories with a scalar sector, T$e$V$e\,$S  evidences \textit{no} cosmological evolution of the Newtonian ``constant'' $G_N$ (Bekenstein and Sagi 2008).   My original choice of $\mathcal{F}(y)$ involves a singular passage into the cosmological domain; more successful choices have been propounded by Zhao and Famaey (2006) and Sanders (2006).

The post-Newtonian behavior of T$e$V$e\,$S in terms of the celebrated parametrized post-Newtonian coefficients has been investigated (Bekenstein 2004a and 2005, Giannios 2005, Sanders 2006, Tamaki 2008).  The $\beta$ and $\gamma$ parameters exactly agree with those of GR.  The parameters $\zeta_1, \zeta_2, \zeta_3, \zeta_4$ and $\alpha_3$  are all expected to vanish here as in GR because 
T$e$V$e\,$S is derived from an action principle, and is thus a conservative theory.  Eva Sagi (unpublished Ph. D. work) has shown that the preferred location parameter $\xi$ vanishes too.  There are strong indications that the preferred frame coefficients $\alpha_1$ and $\alpha_2$ are non-vanishing, but extant calculations of them are in want of consistency.  Reliable calculation of these two coefficients is a high priority because experimental bounds on them are strong.

Out of the parameters of T$e$V$e\,$S one can construct the MOND scale of acceleration.  Omitting a factor close to unity and reinstating $c$ this is
\begin{equation}
a_0=\frac{c^2 \surd k}{4\pi \ell}\ .
\end{equation}
The Newtonian regime mentioned above sets in when the naive gravitational field is well above $a_0$.  For weaker ones which are not far from spherically symmetric, the gravitational field obeys an equation like~(\ref{AQUAL}) with $\tilde\mu$ directly related to the T$e$V$e\,$S $\mu$ function (Bekenstein 2004a, Bekenstein and Sagi 2008).  Our stipulation of the small argument limit of $\mathcal{F}$ leads automatically to the extreme MOND limit
for fields $\ll a_0$.  With a suitable form of $\mathcal{F}$ at intermediate arguments, T$e$V$e\,$S thus reproduces the essence of AQUAL and MOND, and inherits their success in the theater of galaxies.
To conclude one should mention that when worked out in detail, the scale $a_0$ is found to have a rather weak  dependence on epoch of the universal expansion (Bekenstein and Sagi 2008). 

Sanders (2005) has proposed a variant of T$e$V$e\,$S, a bi-scalar tensor vector theory (BSTV) with three free functions and a free parameter.   BSTV  is a more appropriate frame for generating cosmological evolution of $a_0$.  Since numerically $a_0\sim cH_0$, it is often argued that this scale must be determined by cosmology, and should thus vary on a Hubble timescale (Milgrom 1983a).  As mentioned,  in T$e$V$e\,$S $a_0$ evolves very slowly, but its change is faster in BSTV.   Discrimination between the two behaviors may be possible with good rotation curves of disk galaxies at redshifts $z= 2$---5.  Such curves are just now coming into range. 

Zlosnik, et al. (2007) have also proposed a variant of T$e$V$e\,$S, a tensor-vector theory in which the timelike vector is normalized with respect to $\tilde g_{\alpha\beta}$.   The vector's action is taken to be a function $F$ of ${\cal K}$, the  quadratic form in the derivatives of the vector field from Einstein-Aether theories.   The theory has four parameters:  a length scale and three dimensionless parameters.  The form of  $F({\cal K})$ can be deduced approximately from the requirement that MOND arise in the nonrelativistic quasistatic limit, and from the stipulation that a cosmology built on this theory shall have an early inflationary period and an accelerated expansion at late times.

\section*{5. Gravitational lenses and cosmology in T$e$V$e\,$S}

How does T$e$V$e\,$S  measure up to the task of
describing gravitational lensing?   Because all matter and field actions, apart of those for $g_{\alpha\beta}, \phi$ and $\mathcal{U_\alpha}$, are constructed with the metric $\tilde g_{\alpha\beta}$,  light rays in T$e$V$e\,$S are null geodesics of  $\tilde g_{\alpha\beta}$, which is built out of  $\mathcal{U_\alpha},  \phi$  and $g_{\alpha\beta}$.   Since the lensing by galaxies and clusters of galaxies spans cosmological distances, one must first understand a little about T$e$V$e\,$S cosmology. In its isotropic   cosmological models the $\mathcal{U_\alpha}$ is pointed precisely in the time direction.  Consequently,  models with baryonic matter content alone tend to be very similar to the corresponding GR models because the scalar $\phi$'s energy density stays small (Bekenstein 2004a, Chiu et al. 2006, Zhao, Bacon et al. 2006) while the vector's is nill.   According to Chiu et al.   these models give a reasonable relation between redshift and angular distances, and provide just as good  a scaffolding  for the analysis of cosmologically distant gravitational lenses as do GR models.  

As in isotropic cosmological models, so in static situations like the environment of a galaxy, the vector $\mathcal{U}^\alpha$ is pointed precisely in the time direction.  To compute the light ray deflection in linearized theory one also needs the scalar field $\phi$ and the metric $g_{\alpha\beta}$, both to first order in the Newtonian gravitational potential $\Phi_N$.  The line element takes the form (Bekenstein 2004a, Bekenstein and Sagi 2008)
\be
d\tilde s^2 =-(1+2\Phi)dt^2+(1-2\Phi)(dx^2+dy^2+dz^2),
\label{le}
\ee
where $\Phi=\Phi_N+\phi$; here we again omit the mentioned nearly unity factor.  (In linearized GR one gets the same form of metric, but with $\Phi_N$ in place of $\Phi$.)
Since the same potential $\Phi$ appears in both terms of this isotropic form of the line element, the light ray bending, which leans on both to equal degree, measures the same gravitational potential as do nonrelativistic dynamics, which are themselves sensitive only to the temporal part of the line element.    Thus in  T$e$V$e\,$S  an extragalactic system lenses light, or radio waves, just as would GR, were the latter supplemented by DM    in the amount and with the distribution necessary to reproduce the observed galactic dynamics (Bekenstein 2004a). 
 
   In GR $\Phi_N$ is all there is, and its Laplacian, as determined from the lensing observations or from the dynamics,  will give the \textit{total}  mass distribution.  But DM is not visible directly, so its mass distribution (inferred by subtracting the observed baryonic component) cannot be checked; its status is better described with the terminology of physical plausibility.   By contrast in  T$e$V$e\,$S the observationally determined $\Phi$ is to be broken up into two parts: $\Phi=\Phi_N+\phi$.   The $\Phi_N$ part is given by Poisson's equation,  the $\phi$ part by a nonlinear AQUAL-type equation (compare \Eq{scalaraction} with \Eq{lagrangian}); the observed  \textit{baryonic} mass density $\rho$ is the source of both.   Evidently the  T$e$V$e\,$S scheme is falsifiable---by comparison of the calculated potential with that inferred from the lensing ---to a larger extent than is the DM paradigm for which any discrepancy can be tucked away into the invisible component.

Some features of  T$e$V$e\,$S  gravitational lensing by a pointlike mass $M$ are worked out by Chiu, et al. (2006).   They note that the deflection angle in the  deep MOND regime [impact parameter $b\gg b_0\equiv  (k/4\pi)(GM/a_0)^{1/2}$] approaches a constant, as might have been expected from naive arguments, but is less  predictable in the intermediate MOND regime $b\sim b_0$.   This may serve as a warning against doing lensing analysis with a mixture of  GR and MOND concepts (e.g. Mortlock and Turner 2001).  In analogy with GR, Chiu, et al. work out the  lens equation for T$e$V$e\,$S  (which controls the amplifications of the various images in strong lensing).  For two images they find that the difference in amplifications is no longer unity, as in GR, and may depend on the masses.   They also investigate the gravitational time delay in T$e$V$e\,$S which is important for interpreting differential time delays in doubly imaged variable quasars.

The baryon distribution in a galaxy is better represented by the Hernquist model than by a point source model.  Zhao et al. (2006) employ both to compare T$e$V$e\,$S  predictions with a large sample of quasars doubly imaged by intervening galaxies.    The lensing galaxy masses are estimated by comparing observations both with predicted image positions and with predicted amplification ratios; the two methods are found to give consistent results, themselves well correlated with the luminosities of the galaxies.  The corresponding mass-to-light ratios are found to be in the normal range for stellar populations, with some exceptions.  This result clashes with the claim by Ferreras et al. (2008) that lensing by galaxies from the very same sample can only be explained in MOND by including a lot of DM apart from neutrinos.  But the last authors use a mixture of MOND and GR instead of  T$e$V$e\,$S.

What should the probability distribution by angular separation of the two images in a sample of lensed quasars?  This important question has proved troublesome for the DM paradigm.  In T$e$V$e\,$S it has been investigated by Chen and Zhao (2006) and lately by Chen (2008).   Again modeling the shapes of the mostly elliptical galaxies with Hernquist profiles, and describing their space distribution with the Fontana function, these workers compare predictions of both T$e$V$e\,$S for a purely baryonic universe with cosmological constant and of GR with DM and baryons with  the CLASS/JVAS  quasar survey.   After the preliminary work the later paper reports that  T$e$V$e\,$S comes out on top.  All the above is accomplished with spherical mass models of the galaxies; a step towards the modeling of asymmetric lenses within  T$e$V$e\,$S has been taken by Shan, Feix et al. (2008).

When it comes to weak lensing (distorted but unsplit images) by clusters of galaxies, a pure MOND account is less than satisfactory.  The case of spherically symmetric clusters is fairly summarized by Takahashi and Chiba (2007).  In spherical symmetry the Poisson and AQUAL equations of weak field T$e$V$e\,$S for $\Phi_N$ and $\phi$ are easily solved, and the total potential $\Phi=\Phi_N+\phi$ is found to be related to $\Phi_N$ by a MOND relation with some complicated $\tilde\mu$ function.    These authors use several such interpolation functions to predict the shear and convergence of the light lensed by a large number of quasispherical clusters in terms of the visible baryonic matter, but fail to get a fit with observations unless they add a neutrino component \textit{a la} Sanders (2003, 2007); the required neutrino mass is unrealistically large, so it seems that a DM component is needed to buttress the MOND effect.   Similar conclusions are reached by Natarajan and Zhao (2008).

Nonspherical cluster systems are also problematic.  In the massive colliding clusters systems  MACSJ0025.4--1222 (Brada\v c et al. 2008) and 1E0657--56 (Clowe et al. 2004) the galaxy components have been rudely separated from the hot gas concentrations.    Weak lensing mapping  using background galaxies shows the gravitating mass to be preponderately located in the regions containing the galaxies, rather than in the gas which accounts for the bulk of the visible baryonic mass (Clowe et al 2006, Brada\v c et al 2008).  Collisionless DM would indeed be expected to move together with the galaxies and get separated from the collisionless gas; hence the widespread inference that much DM exists in these systems.  However, this view conflicts with the finding (Mahdavi et al. 2007) that in the merging clusters A520 the lensing center is in the hot gas which is separate from the galaxy concentration. Angus, Famaey et al.  (2006)  considered it possible to explain the lensing seen in 1E0657--56 by T$e$V$e\,$S with  a reasonable purely baryonic matter distribution, but later concluded (Angus, Shan et al. 2007) that a collisionless component is needed after all, with neutrinos just barely supplying a resolution.   This conclusion is confirmed by a careful study of Feix, Fedeli et al. (2008) who devised a Fourier solver for the AQUAL equation in T$e$V$e\,$S, and conclude that the source of gravity in 1E0657--56 must include an invisible component.  This study also shows that nonlinearity of the AQUAL equation cannot prevent the lensing from tracking the baryonic matter.

The weak lensing by cluster Cl0024+17 provides another relevant case study.  Jee et al. (2007) find its deduced mass surface density to exhibit a ring which does not coincide either with the galaxy distribution, or the hot gas.  Again this has been hailed as graphic proof of DM.  But Milgrom and Sanders (2008) argue that such feature is actually expected in MOND, lying as it does at the transition between the Newtonian and the MOND regime.  Famaey et al. (2007) conclude that the lensing in  Cl0024+17 can be modeled in MOND by including 2 eV neutrinos.  A truly T$e$V$e\,$S model of  Cl0024+17 is still outstanding.

Turn now to cosmology.  Critics of MOND used to argue that the complex power spectrum of cosmological perturbations of the background radiation, which is said to  be well fit by the ``concordance'' DM model of the universe,  proves that DM is essential to any rational picture of the cosmos.  They tacitly assumed that the MOND paradigm could never measure up to the same test.  With T$e$V$e\,$S on the scene this could be checked for the first time.
 
 In a massive work Skordis (2006) has provided the full covariant formalism for evolution of cosmological perturbations in T$e$V$e\,$S, the analog of that in GR.  Using it Skordis et al. (2006) have shown that, without invoking DM, T$e$V$e\,$S can largely be made consistent with the observed spectrum of the spatial distribution of galaxies, and of the cosmic microwave radiation, if one allows for contributions to the energy density of massive neutrinos, and of a cosmological constant.  The role of DM in GR cosmology is taken over by a feedback mechanism involving the scalar field perturbations.  Dodelson and Liguori (2006) have independently calculated perturbation growth, and claimed that it is rather the vector field in T$e$V$e\,$S which is responsible for growth of large scale structure without needing DM for this.  Thus although the elimination of galaxy bound DM was the original motivation for MOND, and thus for T$e$V$e\,$S,  the later may potentially provide a way to eliminate cosmological (homogeneously distributed) DM !
 
Apart from the DM mystery, cosmology furnishes us with a  ``dark energy'' mystery.  Dark energy is the agent responsible for the observed acceleration of the Hubble expansion in the context of GR cosmological models.  An interesting question is whether modified gravity can supplant the dark energy. Here we touch only upon T$e$V$e\,$S related work in this direction.    Diaz-Rivera et al. (2006) find an exact deSitter solution of T$e$V$e\,$S cosmology which can represent either early time inflation epochs or the late time acceleration era.   Hao and Akhoury (2005) conclude that with a suitable choice of the T$e$V$e\,$S function $\mathcal{F}$, the scalar field can play the role of dark energy.   According to Zhao (2006)  the choice of $\mathcal{F}$ implicit in the work of Zhao and Famaey (2006) leads to cosmological models that evolve at early times like those of standard cold DM  cosmology, and display late time acceleration with the correct present Hubble scale, all this without needing DM or dark energy. Likewise, in the related Einstein-Aether theory, Zlosnik et al. (2007) find that with suitable choice of  their theory's ${\cal F}$, the vector field can both drive early inflation as well as double for dark energy at late times. 

GR and T$e$V$e\,$S differ, also in the cosmological arena, in the relation they stipulate between the matter overdensity and the local depth of the gravitational potential.  The advent of large lensing surveys may open a way to distinguish between these two theories, as well as between GR and other modified gravities, by exposing correlations between galaxy number density and weak lensing shear  (Schmidt 2008, Zhang et al. 2007).  The effect of the dark energy on the expansion can be separated out by comparing cosmological models with the same expansion history in two theories.   And the ultimate confrontation between GR and T$e$V$e\,$S cosmology may be accomplished by cross-correlating galaxy number density with cosmic microwave background antenna temperature (Schmidt et al. 2007).

\end{document}